\documentclass[preprints,review,accept,oneauthor,pdftex]{Definitions/mdpi} 

\usepackage{float}

\usepackage{amsthm}
\usepackage{amssymb}
\newcommand{\be}{\begin{equation}}
\newcommand{\ee}{\end{equation}}
\newcommand{\bea}{\begin{eqnarray}}
\newcommand{\eea}{\end{eqnarray}}

\firstpage{1} 
\makeatletter

\newcommand{\msun}{\, {\rm M_{\odot}}}

\makeatother
\pubvolume{1}
\issuenum{1}
\articlenumber{0}
\doinum{}
\pubyear{2022}
\copyrightyear{2022}
\history{}
\pdfoutput=1

\usepackage{url}
\usepackage{hyperref}

\Title{Constraints on PBH as dark matter from observations: a review}

\Author{Marc Oncins $^{1,2,3}$ \orcidA{}}

\address{%
$^{1}$ \quad Institut de Ciències del Cosmos, Universitat de Barcelona, Martí i Franquès 1, 08028 Barcelona, Spain. \\
$^{2}$ \quad Departament de F\'isica Qu\`antica i Astrof\'isica, Facultat de F\'isica, Universitat de Barcelona, 
Mart\'i i Franqu\`es 1, 08028 Barcelona, Spain \\
$^{3}$ \quad Institut d'Estudis Espacials de Catalunya (IEEC), Edifici Nexus I, C/ Gran Capit\`a, 2-4, desp. 201, 08034 Barcelona, Spain.}

\corres{Correspondence: oncins@icc.ub.edu}

\abstract{Primordial black holes (PBHs) are a fascinating candidate for being the dark matter, albeit one which has been heavily constrained. This review presents an in depth look at those observational constraints, particularly at their nuances and uncertainties. Despite their varied origins, the standard PBH formation path is assumed to be collapse of perturbations after inflation, which should leave signals visible in the CMB at certain scales. Other constraints come from microlensing surveys, which severely limit PBHs as dark matter in the solar to satellite range, but there are diminishing results in regards to lower mass ranges. Gravitational waves signals and PBH evaporation from Hawking radiation also make for useful probes, but the former requires the next generation of experiments before making constraints beyond the solar mass range, and the later is severely limited above $10^{-16} \msun$. Other dynamical and accretion constraints exist for PBH of large masses. Care also has to be given, as all these constraints can carry different implications coming from differences between monochromatic and extended mass distributions, and their degree of clustering. Beyond all these issues, a window still exists for primordial black holes to be all of the dark matter between $10^{-16}$ and $10^{-11} \msun$.
}


\begin{document}


\section{Introduction}\label{sec:intro}

Primordial black holes (PBHs) are one of the current most interesting candidates to be the dark matter (DM). Black holes (BHs) emerging from the very early universe, without any of the restrictions on mass or abundance that characterize BHs of stellar origin. Not only they fulfill the basic criteria needed for the DM, they do so without the need to invoke a new set of existing particles. Furthermore, while their specific mass is not set in stone, on cosmological scales PBHs would behave like the standard particle cold dark matter \cite{2021arXiv211002821C}.

While this might make them look ideal, the formation of PBHs is not so simple and often requires some form of new physics too, so their specific origins and abundance are still hotly debated. While first theorized in \cite{1967SvA....10..602Z, 1971MNRAS.152...75H}, their name and characteristics, including formation from from primordial inhomogeinities during inflation and their cosmological implications, were first formalized in \cite{1974MNRAS.168..399C, 1975Natur.253..251C, 1975ApJ...201....1C}. Today a variety of formation paths exist, including exotic ones. For a review of the most standard formation path and possible alternatives I refer the reader to \cite{2022Univ....8...66E}.

PBHs as a DM candidate have received considerable attention since the LIGO \cite{2015CQGra..32g4001L} discovery of gravitational waves \cite{2016PhRvL.116f1102A} and its possible attribution to PBHs \cite{2016PhRvL.116t1301B}. Lately, they have been mostly ruled out of originating all the gravitational waves detected by the LIGO collaboration \cite{2016PhRvL.117f1101S, 2016arXiv161008725W, 2021PhRvD.103b3026W}, though some doubts still remain \cite{2020JCAP...09..022J}.
PBHs however can form in a wide mass range, depending only on the circumstances of the collapse that forms them. Despite this very wide range of parameters, PBHs being all of the DM is strongly constrained in most other mass ranges through a rather diverse set of observations \cite{2021arXiv211002821C, 2018CQGra..35f3001S,2021RPPh...84k6902C,2020PhRvD.101f3005S}. While the number of possibilities makes them a very exciting prospect from both a theoretical and observational standpoints, it also means it is very hard to track the extent to in which masses the PBHs can be all of the DM or more particularly the nuances and uncertainties associated with these constraints.

In this review I attempt to make a brief summary of all current known constraints on PBHs as DM, but also put special focus on the physical origin of the constraints and the implications they carry for both their current form and future prospects. A nuanced understanding of both is helpful when the constraints appear to constantly be changing, and in correctly assessing the possibilities of future detection. Previous review with extensive reference lists can be found in \cite{2021arXiv211002821C, 2018CQGra..35f3001S}.

In section \ref{sec:Form} I make a very brief summary of what is usually considered the standard formation paths and what constraints can be derived from it. The following section \ref{sc:Mono} will be an overview of all known current monochromatic constraints, divided by their types. Section \ref{sc:non-diff} will tackle the issue of how non-monochromatic constraints work and their issues, and also how clustered PBH distributions can become relevant and modify existing constraints. Finally, I will detail the still existing window on PBHs as DM and works that have tried to close it in \ref{sc:window}, the possibility of directly detecting the PBHs in section \ref{sec:Direct detection} and present my conclusions and future prospects in section \ref{sec:future_perspectives}.

\section{Formation}\label{sec:Form}

As this review's main purpose is to detail the existing observational constraints on PBHs as DM, I will keep explanation on PBHs possible origin brief.

The most common origin for PBHs is in the very early Universe, during radiation domination, where large curvature perturbations generated during inflation could have undergone gravitational collapse. This early epoch should be radiation dominated, so our focus in on energy density, which we will call it $\mu$. For the case of Planck units, which take $G=c=1$, the Schwarzschild radius of a BH can be simplified from the standard form $r_s = 2GM/c^2$ to $r_S = 2M$.

It is also obvious that the total mass within any spherical symmetric perturbation in Planck units would be $M\sim \frac{4}{3}\pi \, \mu r^3$, as $c=1$. Using that with the simplified Schwarzschild radius definition means the condition for the perturbations to collapse is \cite{1974MNRAS.168..399C}:

\begin{align}
	r_S = 2\frac{4\pi}{3} r_S^{3}\mu \longrightarrow \mu r^2 \gtrsim 1 \;,
	\label{eq:PBhform}
\end{align}
where we drop the constants on the right term. While this condition might appear strange, it is in fact very close to the Jeans length for oscillations, which in natural units would be $\lambda_J \sim \left(\frac{\pi}{\mu}\right)^{1/2}$. Another possibility is simply to remember that in radiation dominated universe the radiation pressure is $p = \frac{\mu}{3}$. Matching it with the gravitational potential to see when the latter wins out and the region will collapse ends giving the same result as above. 

A density contrast can then be introduced, which compares the energy density of a point $\mu$ with the background density $\mu_0$. We will call this contrast $\delta$ and use the following definition:
\begin{align}
	\delta = \frac{\mu-\mu_0}{\mu_0} \;.
	\label{eq:delta}
\end{align}

This allows us to measure how overdense or underdense a certain region is, and so the strength of the perturbation. In an homogeneous, isotropic and flat universe, given condition \ref{eq:PBhform}, this will end requiring $\delta \gtrsim 1$ for the perturbations to collapse \cite{2018CQGra..35f3001S}.

This condition is the threshold: if $\delta$ fulfills it by being greater than the threshold, the perturbation will collapse and form a PBH. If it is lower the pressure of radiation will win out, resulting in the perturbation continuing its spread maybe even until today, though it will so be heavily redshifted as to be extremely hard to detect. 

This is all for a very idealized case though. From the start the delta will be a function rather than a uniform value. Numerical simulations have found that the exact threshold does not depend only on the maximum of the perturbation, but also on the shape of the $\delta$ function \cite{1999PhRvD..59l4013N, 1999PhRvD..60h4002S, 2005CQGra..22.1405M,2015PhRvD..91h4057H, 2019PhRvD.100l3524M, 2020PhRvD.101d4022E}. Thus for different perturbations the threshold will differ even if their maximum and average are the same, unlike our estimation. 
 
Those are not the only issues, as the definition of the threshold has to be gauge independent and take into account different equations of state of the universe, to account for all possible inflation models. A matter dominated universe or a intermediate, dust like, universe for example would change the physics of the collapse. All of these examples can radically alter the threshold, which makes a generalized form to compute the collapse a necessity.

A way to improve upon our definitions is to use a new and more general type of condition. Called the critical threshold $\delta_c$, it is not a direct measure of the over-density like $\delta$ but rather of the mass excess per a unit of volume, the so called compaction function \cite{1999PhRvD..60h4002S}. It not only is completely generic, the critical threshold also accepts a very good analytical approximation with parameters describing shape and space-time \cite{2020PhRvD.101d4022E, 2021JCAP...01..030E}, which simplifies simulations enormously. Previous analytic estimates of the threshold were also done in \cite{1975ApJ...201....1C, 2013PhRvD..88h4051H}

Despite this, however there is another major complication that brings our first constraint. The current power spectrum expected for our universe should be very close to scale invariant, and at the scales of the power spectrum present in the CMB, the fluctuations are of the order of the microKelvin, $\delta \sim 10^{-5}$ \cite{2020A&A...641A..10P}. Therefore, remembering the condition above the collapse of perturbations into PBHs should not occur in any major way, not enough for them to be the DM. 

There are a variety different solutions to this conundrum. First, some kind of enhancement of the perturbations at only certain scales could have happened. There are in fact a large number of ways to get this kind of enhancement \cite{2017PDU....18...47G, 2018PhRvD..98l3514K, 2019PhRvL.122n1302G}, from slightly modified inflation models to new physics coming from string theory, but all have in common that they are outside the standard model. The most common way is to have a modified inflation potential, the so called ultra-slow roll potential \cite{2017PDU....18....6G, 2019JCAP...09..073A}, which results in perturbations at a certain parameter dependent scale being enhanced up to $\delta \sim 1$ for the isotropic flat case, while remaining the usual value at other scales.

Another related possibility is that the spectrum of fluctuations is non-gaussian. The fluctuations in the CMB are $\delta \sim 10^{-5}$ as measured on average, which is too small for any kind of noticeable PBHs to form if the fluctuations follow a gaussian spectrum. Gaussian fluctuations would be the prediction of the standard model. But if fluctuations do not follow that gaussian spectrum because we have new physics, then there could be a tail of extreme results with $\delta> \delta_c$ without altering the perturbations general properties. This could result in an important number of PBHs forming \cite{1997PhRvD..55.7423B,2018JCAP...03..016F,2019PDU....24..275A}. Currently, in the CMB there has not been any sign of any type of non-Gaussian fluctuations \cite{2020A&A...641A...9P, 2022arXiv220401781C}, which is why a lot of models do not contemplate them in a major way, but non-Gaussian fluctuations are still not completely ruled out \cite{2022arXiv220308232C}.

Something necessary to understand however is that even if we are using a model that results in the enhancing of (gaussian) fluctuations to obtain PBHs, non-gaussianities will still be relevant. This is due to the relation between the $\delta_c$ and the curvature perturbation being non-lineal \cite{2019JCAP...07..048D, 2019PhRvD..99l3501K}, resulting in non gaussianities in the density fluctuations despite a gaussian curvature perturbation. As the abundance of PBHs is very sensitive to the values of $\delta$, any kind of higher than expected value like the ones that would come from an unaccounted tail of fluctuations, would modify the number of PBHs in a major way. This adds another layer of subtlety in computing the abundance of PBHs from a particular model, though workarounds exist \cite{2019PDU....24..275A}. For more studies of non-gaussianities in particular look \cite{2013JCAP...08..052Y, 2019JCAP...09..033Y, 2021JCAP...10..053K, 2022JCAP...05..012E}.

Another possibility is to preserve the standard inflationary model with scale invariant power spectra as it is on the CMB. Even then, PBH formation is possible by creating high subhorizon scale $\delta$ due to some phase transition that may naturally take place as the universe cools down. Most notably, PBHs could be formed in the epoch of QCD phase transition \cite{PhysRevD.55.R5871,Byrnes_2018,2019arXiv190411482G} as there is temporary drop in the pressure that counteracts gravity, allowing for higher $\delta$ than would be expected from the amplitude of the perturbations. PBHs formed in this epoch should have a mass of around the solar mass \cite{2019arXiv190411482G}, though using different phase transitions or other non-standard physics can change the mass range.

Even after all that, that is but the first step of PBH formation. Collapse into a BH does not happen immediately once the perturbation is over the threshold, it is not until the perturbation crosses the horizon that it can collapse, as before the perturbation is not fully causally connected due to inflation. While this might look like just a timing issue it carries important consequences. As the perturbation will collapse in the horizon in which it has reentered, the mass of the resulting PBH will be of the order of the mass of the horizon \cite{1998PhRvL..80.5481N, 2005CQGra..22.1405M, 2021JCAP...05..066E, 2020PDU....2700466E}. The lower the scale, the bigger the horizon and the higher the mass it will have, and vice versa.The mass within the horizon can be estimated with\cite{2016PhRvD..94h3504C}:

\begin{align}
	M_h \sim  \rm \frac{c\,t^3}{G} \sim 10^{15} g \left(\frac{t}{10^{-23}s}\right)  \;,
\end{align}
with the time $\rm t$ ranging from the Planck time ($10^{-43} s$) to much later in the universe's life. This is what gives PBHs such enormous mass ranges. A perturbation generated in first instants of inflation can create tiny BHs that could could explain the whole of the dark matter, while perturbations generated later can result in massive PBHs bigger than any current stellar BHs and which could be the seeds of Supermassive black holes.

Of course, the perturbations that make PBHs do not have any reason to all come from the same exact instant. This will naturally result in a spread of masses for our PBHs. For the case of enhanced perturbations we do expect the enhancement to only happen at certain scales, so while PBHs will have an extended mass distribution, if there are enough for them to be a noticeable part of our universe we also anticipate them to be around a certain peak. Even then this is not a completely hard rule, as there can be extreme cases like double inflation models \cite{2018PhRvD..97d3514I} and also models where PBH form in a completely different way that results in a more even spread \cite{2019arXiv190608217C}. There are still even other relatively exotic methods of PBH formation, such as inhomogeneous baryogenesis \cite{2019JCAP...01..027H}, domain walls \cite{2017JCAP...04..050D,2019EPJC...79..246B}, vacuum bubbles \cite{2016JCAP...02..064G, 2017JCAP...12..044D}, axion driven inflation \cite{2016JCAP...12..031G, 2017JCAP...07..048D} and through non-topological solitons known as Q-balls \cite{1986744}, each of which will have different results for the PBH extended mass distribution.

Another expectation would be that they would form with spin 0 or close \cite{2017PTEP.2017h3E01C}, as they form in the radiation dominated epoch where the combined fluid of radiation and matter sheds angular momentum almost instantly. This can be a clear difference from stellar origin BHs, which will inherit at least a fraction of the spin of their forming star\footnote{How much is a point of contention, with very efficient angular moment transport models positing very low spin\cite{Fuller_2019} while others using medium efficiency average notable spin BHs\cite{2012A&A...537A.146E}. Observations of X-ray binaries always have a very high spin, but they are subset of BHs and so there could an intrinsic bias too.}, though again in more exotic formation paths there could be PBHs with noticeable spin. A good example is the case where PBHs would form in a matter dominated epoch, which should result in PBH having all very high spin instead of close to 0 \cite{2017PhRvD..96h3517H}. As spin can be relevant for a subset of constraints, this is a rather important distinction.

Something similar applies to the degree of clustering with which with PBHs will form. In the simplest case  the PBHs should follow a standard poissonian matter distribution \cite{2003ApJ...594L..71A}, as the perturbations they form from should also come from a homogeneous universe and the enhancement should not a priori have any specific spatial distribution. Exotic models exist however where this is not the case, in particular for alternative origins like the previously mentioned domain walls\cite{2019EPJC...79..246B}. This adds still another layer of complexity and will again affect constraints on their abundance.

Still, as mentioned, if PBHs are formed through perturbation collapse not from phase transition, then the perturbations will also be reflected on the CMB, concretely its temperature angular power spectrum and CMB spectral distortions. Only a number of perturbations will collapse, so the rest will transmit and appear like the other CMB fluctuations. Unfortunately, either through Planck \cite{2020A&A...641A..10P} or other measurements \cite{10.1111/j.1365-2966.2011.18245.x, 1996ApJ...473..576F}, the CMB can only constrain the power spectrum on the lower scales, equivalent to the highest masses of PBHs. Currently, no non-gaussianity nor enhanced fluctuation has been found, limiting these lower scales, but only the lower scales. Saying it in another way, any PBH formation below $10^3 \msun$ in this standard pathway is currently still possible.

\section{Monochromatic constraints}\label{sc:Mono}

PBHs could be the DM, but the properties of these PBHs can be drastically different between different mass ranges. $10^3 \msun$ BHs would gravitationally affect any star or planet that is close to them, while much more smaller $10^{-10} \msun$ BHs could cross the Solar system millions of times and we would not notice them \cite{2009ApJ...705..659A}.

Therefore is common sense to divide the types of PBH by their mass. While extended mass distributions are expected for PBHs, for the standard inflation perturbation collapse case the enhancement at certain scales should result in a particular mass range corresponding to those scales, with a peak. A very helpful approximation with setting constraints is taking the peak of that mass distribution and making as if all the PBHs in it had that particular mass. This simplifies the physics in most cases enormously, as we can avoid integrals through all the masses and streamline the calculations needed to see if it is even worth to pursue certain constraints. And while the final result is an estimate, that estimate should be a fairly close and conservative approximation to the real case of the extended mass distribution, as the peak of that same mass distribution will likely have an abundance only slightly below 1 if it contains all of the DM after all.

\begin{figure}
	\includegraphics[width=\columnwidth]{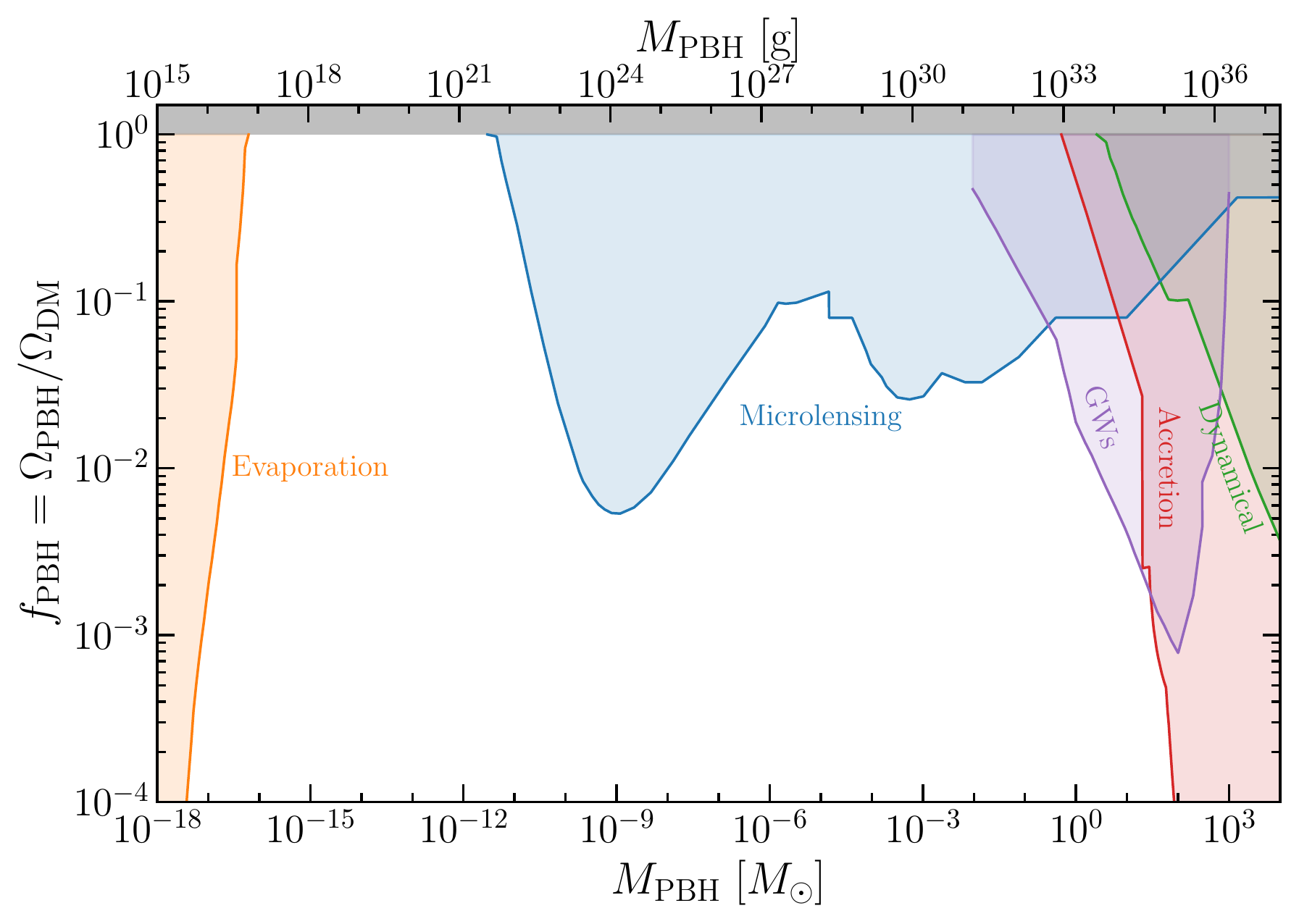}
	\caption{Plot with current existing robust monochromatic PBH constraints, with each color and name corresponding to a different type. The abundance on the left axis is which fraction of the dark matter they could be, while mass is given in both $\msun$ and grams. Evaporation constraints come from \cite{2021MNRAS.504.5475K,2019PhRvL.122d1104B, 2020PhRvD.101l3514L, 2012PhRvD..86d3001B, 2010PhRvD..81j4019C, 2019PhRvL.123y1101L}, microlensing constraints come mostly from a number of surveys, including Subaru, Kepler, MACHO and EROS \cite{2019NatAs...3..524N, 2000ApJ...542..281A, 2014ApJ...786..158G, 2007A&A...469..387T,2022arXiv220213819B,2018PhRvD..97b3518O}, gravitational waves bounds come from LIGO and related studies \cite{2019PhRvD.100b4017A, 2021PhRvL.127o1101N, 2020JCAP...08..039C, 2018PhRvD..98b3536K} and dynamical \cite{2014ApJ...790..159M, 2019PhRvD..99l3023L, 2020PhRvD.101f3019W} and accretion \cite{2019JCAP...06..026M, 2021ApJ...908L..23L} constraints from a variety of sources. More details on how the constraints are obtained are in each of the corresponding subsections. Figure obtained from \cite{PBHBounds}, where all references, other bounds and plotting codes are available.}
	\label{fig:PBH_bounds}
\end{figure}

The end result of this simplifications are monochromatic PBH bounds, which are shown in figure \ref{fig:PBH_bounds}. Of further note is also that the graph is logarithmic and covers more than 22 orders of magnitude in mass. In this way it is only thanks to that approximation that covering all these possible ranges with different types of constraints is feasible.

There are as varied constraints as there are mass ranges, but as seen above there are 4 that dominate. Microlensing is the first one, perhaps the most relevant in terms pf width of the constraints. Following from the previous MACHO experiments \cite{2000ApJ...542..281A}, a large number of PBHs would also result in a large number of detectable microlensing events. Lack of such events can be used therefore to set constraints. Another is BHs evaporation, as the expected result of BHs emitting Hawking radiation \cite{1974Natur.248...30H}. The smaller BHs will have significant high energy emissions, so their observation or more particularly lack of thereof can be used to set abundance constraints on PBHs. Gravitational waves (GWs) are perhaps among the most popular due to the LIGO merger detection \cite{2016PhRvL.116f1102A}. If PBHs were all of the DM in masses within LIGO sensitivity range, then there would be an increased number of mergers over purely stellar origins. The lack of such mergers results in strong bounds for the LIGO mass range.
Last are the dynamical and accretion constraints, only possible on very massive black holes. There are multiple sources for these constraints, but they all share that they focus on the large mass of these black holes and how they would have a large effect on their surrounding which would significantly affect galactic dynamics.


\subsection{Microlensing}

Microlensing is the name given to gravitational lensing on a very small scale. Gravitational lensing itself is the bending of light when a body projects a strong enough gravitational force. In typical optical parlance, the gravity of the body acts as a lens to the light coming from behind it towards us. The most illustrative case is that of the Einstein ring \cite{1936Sci....84..506E}, also known as Chwolson ring \cite{1924AN....221..329C}. A source emits light, and in our line of sight towards the light goes through the lens. For the sake of simplicity we shall assume the lens is a black hole: then the light in the direct line of sight will not go through the BH, but affected by gravity will bend around the lens, eventually reaching Earth from multiple directions. This will results in the viewer seeing multiple sources equidistant to where the source actually is, with the direct line of sight blocked in the case the lens is a BH. In the 3 dimensional case, the light coming from the multiple directions result in a ring surrounding the source, thus the name of ring.

\begin{figure}
	\includegraphics[width=\columnwidth]{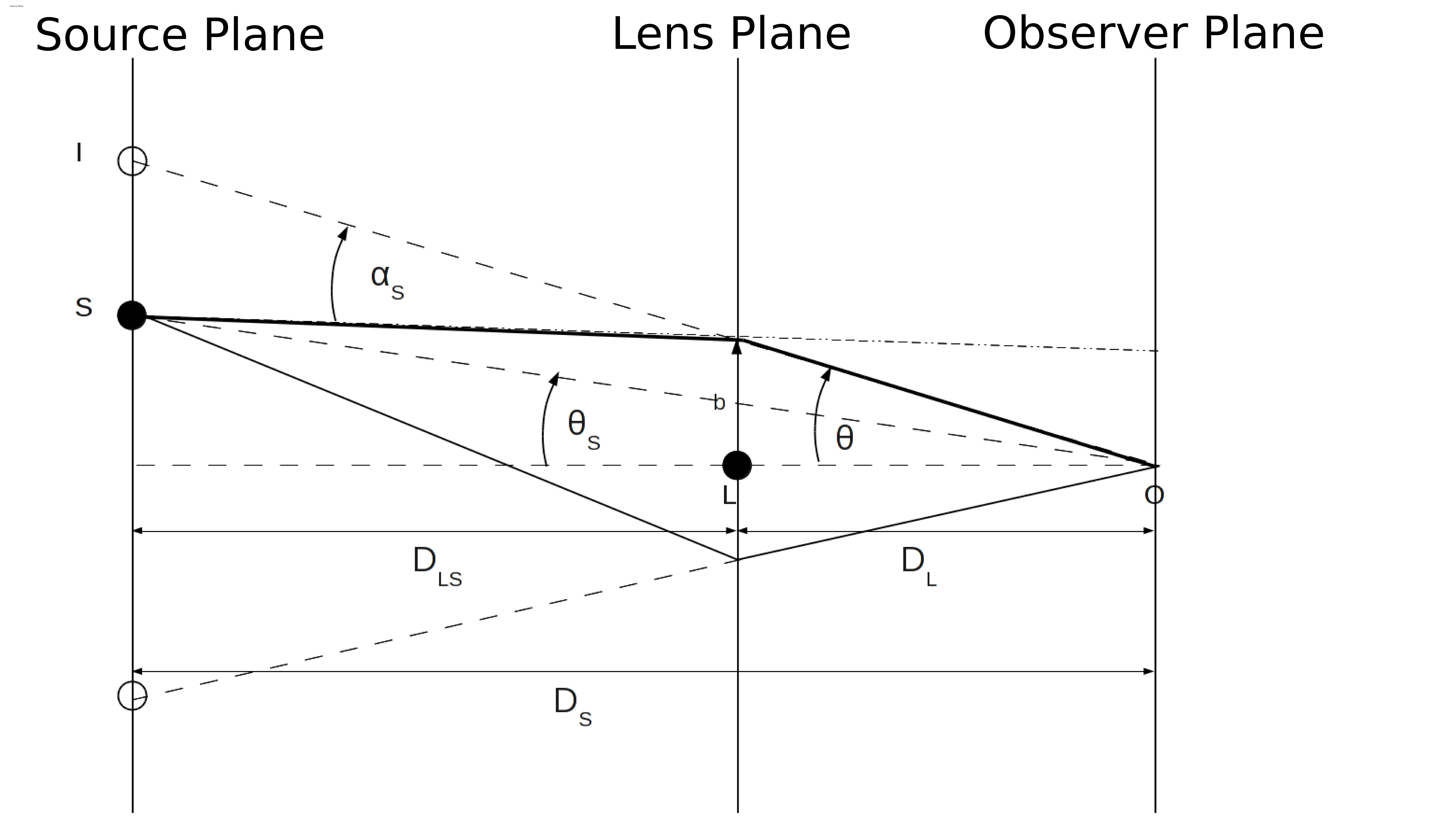}
	\caption{Example of the geometry of a lensing event for a general case. Taken from \cite{universe2010006} Figure 1, $\alpha_s$ is the same as our $\alpha$}
	\label{fig:Lens}
\end{figure}

The Einstein ring is an ideal case, but most lensing events are likely to differ. A generic example for the geometry of a lensing event in shown in figure \ref{fig:Lens}. One particular example is where the object and lens are so far away, or small, that our resolution is unable to separate the two images. This is the specific case of microlensing \cite{1986ApJ...304....1P}, where the lensed image appears all on the same pixel rather than as two or more different images. As we cannot perceive a position shift the typical measurements of angles for each individual image cannot be done, and it can be hard to even notice if the image has been lensed.

However, the presence of more ray beams than expected in a single pixel also means that the image will appear brighter than normal. Going from the figure above, if the rays have similar brightness receiving two or more will make the pixel appear at a higher luminosity than if it only received a single one. Because all stars and galaxies appear to move for an observer on Earth (a combination of both their intrinsic orbits and the Earth's own movement), the lensing event will only happen for the short period of time in which both source and lens are aligned. The increased luminosity can then be compared with the previous one measured just before, and observers can notice and record these kind of changes as lensing events.

To compute the expected magnification of light, a first approximation is to treat gravitational lensing like common optic lenses, using what is called geometric optics. First of all is the deflection angle $\alpha$, which would be how much the ray is deflected from its path \cite{universe2010006} :

\begin{align}
	\alpha = \frac{4GM}{b c^2} \;,
	\label{eq:deflectangl}
\end{align}
where b is the impact parameter, the shortest distance between the point-like lens and the ray it is deflecting, M is the mass of the lens, G the gravitational constant and c the speed of light.

This also using what is called the small angle approximation, $ sin \alpha \approx \alpha$, which works for small $\alpha$. (\ref{eq:deflectangl}) comes from general relativity but the result is a very simple term that reminds of newtonian gravity, so it is also referred as Newtonian approximation. In fact, if we assumed Newtonian gravity and that light is a particle with mass the end result would be exactly half the value of the $\alpha$ in \ref{eq:deflectangl}. Another comparison can be with the Schwarzschild radius of a BH, which can be related by $\alpha = 2 r_s/b$. 

Another parameter of interest is the the Einstein angular radius $\theta_1$, which is the angle at which we would see the previously mentioned Einstein ring. When we are not in the ideal case like in the figure above, $\theta_1$ does differ from the angle at which we see the image $\theta$. Using trigonometry, $\theta_1$ can be computed from equation (\ref{eq:deflectangl}) in the ideal case, the definition of which will carry on for more complex geometries:

\begin{align}
	\theta_1 = \left(\frac{4GM}{c^2} \frac{D_{LS}}{D_s D_l} \right)^{1/2} \;,
	\label{eq:Einst_angl}
\end{align}
where like before there is the implicit assumption that $sin\theta_1 \sim \theta_1$. In standard lensing this can troubling, but in microlensing where the angles are so small it always applies. For PBHs in particular there is again an easy correspondence with the Schwarzschild radius, $\theta_1 = \sqrt{2 R_{Schw}D_{LS}/(D_s D_l)}$

Even in the non-ideal case where $\theta \neq \theta_1$ the Einstein angular radius can be understood as the effective radius in which the the lens bends the light. The timescale of lensing will be therefore be the time it takes the lens to cross $2 \theta_1$ from the Earth's point of view. As the Einstein radius has a dependence on mass $M^{1/2}$, the time scale will become increasingly shorter for the lower mass PBHs, but the exact timescale has heavy dependence on the location of the event with respect to the observer. 

Looking at other important angles in figure \ref{fig:Lens}, $\theta_S$ will be the angle of the source with respect to the observer, and $\theta$ the angle the lensed images would appear at. In the geometric optics approximation the angles $\theta$ and $\theta_S$ can be related with the Einstein radius $\theta_1$ by applying the lens equation of geometric optics to our case \cite{10.1143/ptp/90.4.753}:

\begin{align}
	\theta-\theta_S = \frac{\theta_ 1^2}{\theta} \;.
	\label{eq:lens}
\end{align}

As $\theta_S$ is a single value, it is easy to see there would be two solutions for $\theta$, corresponding to two images for the 2-dimensional case. In the case of microlensing we are unable to distinguish them, but we will be able to distinguish the increasing flux, the magnification of the source the 2 images will cause. 

The two images can be related to the flux by understanding that the flux (f [erg/sec]) results from integrating the surface brightness of the object (S [erg/sec/sr]) for a particular solid angle $\Omega$, the width of the beam so to speak. This solid angle depends on the angle it reaches the observer. Because for microlensing the angles are very small, this integral can be approximated as a simple multiplication $f = S d\Omega$, and compared with the case where there would be no lensing:

\begin{align}
	A=\frac{f_{obs}}{f_0}=\frac{S d\Omega (\theta)}{S d\Omega(\theta_S)} = \frac{d\Omega (\theta)}{ d\Omega(\theta_S)} = \frac{\theta d\theta}{\theta_S d\theta_S} \;.
\end{align}

Computing the magnification then is relatively straightforward if we remember the lens equation again (\ref{eq:lens}):

\begin{align}
	A=\frac{\left(\frac{\theta_S}{\theta_1}\right)^2 +2}{2\left(\frac{\theta_S}{\theta_1}\right)^2
		\sqrt{\left(\frac{\theta_S}{\theta_1}\right)^2 +4}} \pm \frac{1}{2} \;.
	\label{eq:A_m}
\end{align}

Searching for this moment of greater emission from the source a lensing event can thus be detected. Despite the previous complications, the final equation as seen above depends only on the source angle and the Einstein angle. By detecting the magnification of light coming from a source many details can be easily obtained from the lens-source system.

This makes microlensing a very attractive prospect for PBH as DM constraints. We know the mass we want to constrain and we can use the DM galaxy mass distribution to estimate the likely places where the PBHs would be and their velocity distribution. With this the expected magnification can be estimated and if it is within the sensitivity of any scientific instruments. The birth of dedicated surveys like the original MACHO\cite{2000ApJ...542..281A}, but also now Subaru \cite{2019NatAs...3..524N} has allowed to map a mass range of 10 orders of magnitude with constraints, as they have found few or none gravitational lensing events, whereas if those PBHs were the dark matter all models predicted many more.

But despite all these successes, microlensing constraints have started struggling to increase constraints in new mass ranges. This comes from a lot of the approximations used previously. First of all, we have used standard geometric optics, but that approximation requires light rays to behave completely classically. If any of the wave.like properties of light are relevant during the lensing, the whole approximation breaks down. Indeed, as the PBHs we are interested in are extreme objects, their Schwarzschild radius is very small and can become of the order of the wavelength of the light we are observing, triggering wave-like effects from the light that need to be accounted for.

Fortunately, even without using geometric optics the magnification can still be computed, though it will be more complex. As we do not expect the rays to be perfectly symmetrical, the rays very likely traverse different distance to reach the viewer, there will be a time delay between them. This time delay will also result on the beams not reaching on the same phase of the wave. The time delay is much harder to compute, but an analytic expression is still possible \cite{2018JCAP...12..005K}:

\begin{align}
	\Delta t = \frac{1}{c}\frac{D_L D_S}{D_{LS}}(1+z_{s})\left(\frac{\left|\theta-\theta_{s}\right|^2}{2} - \Psi(\theta) \right) \;,
	\label{eq:Delta_t}
\end{align}
where all parameters are the same as previously, $z_s$ is the redshift of the source and $\Psi$ is the so called lensing potential. This lensing potential is related to the density of the lens by the Poisson equation, and can be thought of as projection of the density through the plane of the lens. For point like sources, like our black hole, $\Psi(\theta)= \theta_{1}^2 log \theta \,$ \footnote{The lensing potential is the result of integrating the density through the entire plane of the lens and includes an indeterminate constant. This is normally not an issue as the time delay is used to compare with the time delay from other rays, but it will not help in computing absolute time delays for a path.}.

Due to wave nature of light, this time delay will create a phase difference between the different rays. As the rays converge, this phase difference can result in both destructive and positive interference. As there are multiple rays and both the lens and source are moving with respect to each other, rather than fully positive or fully negative there will be instead interference fringes. This will make the magnification equation grow more complex \cite{2018JCAP...12..005K}:

\begin{align}
	A = \left( \frac{W}{2\pi i} \int d^2x e^{i\omega\Delta t(x,\theta_S)} \right)^2 \;.
	\label{eq:A_inter}
\end{align}

We can no longer avoid an integral or the use of imaginary numbers. W is a dimensionless frequency, related to the frequency of the light $\omega$ and the time delay $\Delta t$. For more details on the complete derivation of $A$ I refer the reader to \cite{2018JCAP...12..005K}.

While these changes might appear minor and too cumbersome, they are relevant because this mistake did in fact happen with Subaru, where an early prepublication version of the paper did not take this into account as noted in \cite{2018PhRvD..97d3514I}. Even with the added complexities, the amplification in equation (\ref{eq:A_inter}) can still be computed to get the corrected constraints as they were finally present in the printed version \cite{2019NatAs...3..524N} and as they appear in figure \ref{fig:PBH_bounds}. The integral makes it harder to compute, but the magnification is still there. Note however that this also makes the constraints much more sensitive to assumptions about the lens and sources used to obtain said magnification. \cite{2020PhRvD.101f3005S} claimed that the constraints obtained by the HSC were overestimated even in the printed version, as they had assumed all the sources from their stellar observations to be of one solar radius, rather than account for the selection bias towards brighter (and so larger) stars.

It is much more problematic however when finite source effects start appearing. These come into play when the source can no longer be taken to be point-like: if the source is big in respect to the lens, emission from different points of the source will be lensed differently. Possibilities include the source not being small enough or far enough, resulting in different time delays between even close by or symmetrical rays and the interference fringes being washed out the larger the effect is. The end result is that the lensing effect disappears, there being no difference with the normal luminosity. Assuming that the emission radius of the source is roughly the same as the radius of the source itself $R_s$, it results in the following condition \cite{2018JCAP...12..005K}:

\begin{align}
 \frac{R_s/D_s}{R_E/D_L} <<1 \;,
\end{align}
where $R_E$ is the Einstein radius in units of distance, $R_E = \theta_1 D_L$. If the condition is not fulfilled, the microlensing effects will start disappearing. Most microlensing surveys of interest are focused on far away objects like M31 for Subaru \cite{2019NatAs...3..524N}, with both lens and source expected to be much closer to each other than Earth. In that case $D_s \sim D_L$, leaving a much simpler condition:
\begin{align}
 \frac{R_s}{R_E} <<1 \;.
\end{align}

Small PBHs have very small Schwarzschild radius and so Einstein radius. This means in effect that distant sources used in lensing, stars from neighbouring galaxies, are simply too big to reliably produce any constraints for PBHs below $10^{-11} \msun$. Therefore either close stars or other very small sources are needed to constrain the mass range. \cite{2018JCAP...12..005K} estimated that we would need a subset of Gamma Ray Burts (GRB) with very small sizes and very small wavelengths, to reliably constrain this mass range but found current GRB detection rates and GRB's intrinsic variability made it very challenging for new constraints to be created.  Other proposals include using pulsar timing arrays \cite{PhysRevD.100.023003} in the radio spectrum for this kind of purpose, as pulsars are smaller than stars, but also reached the conclusion it would take a future and very ambitious observatory to reach mass ranges which are not currently constrained. The use of microlensing in X-ray pulsars however is much more promising, even if current instruments cannot constrain PBHs as DM \cite{2019PhRvD..99l3019B}. 

Other ideas include using Imaging Atmospheric Cherenkov telescopes to detect the extremely fast (t<1s) microlensing events expected for lower asteroid mass range, but prospects are also pessimistic \cite{2022icrc.confE.495P}. Another possibility would be the use of GRB lensing parallax rather than standard microlensing. In a lensing parallax we observe the same GRB via two spatially separated instruments, if the GRB is lensed and the separation of the instruments is of the order of the Einstein radius of the lensing system, then each of the observers will see a different magnification than the other. Asteroid mass PBHs can be constrained if we observe enough GRBs with this technique \cite{1995ApJ...452L.111N, 2020PhRvR...2a3113J}, but the large separations needed between the instruments (of the order of 1 Astronomical unit) means it would require brand new spacecraft detectors.

Microlensing can still remain useful for PBHs of high mass, as recent bounds using quasars showcase \cite{2022arXiv220304777E}, but the much lower number of PBHs means much lower expected number of events, increasing the observation time required for adequate constraints, on top of low mass PBHs being much more interesting for the purpose of dark matter as figure \ref{fig:PBH_bounds} illustrates. 

In summary, while extremely useful, microlensing seems to have reached the limit in what it can constraint, or at least reached very diminishing results. While planet mass PBHs cannot be all of the dark matter, asteroid mass PBHs are free from such constraints.

\subsection{Black hole evaporation} \label{PBH_evap}

PBH evaporation is a process resulting from Hawking radiation emission \cite{1974Natur.248...30H}. Hawking radiation itself is the conclusion of applying quantum effects to a BH's horizon and results in the emission, mostly in terms of particles, of energy depending on the effective temperature of the Black Hole. This emitted radiation would have an almost thermal black body spectrum of the corresponding effective temperature, and this temperature depends mostly on the mass. The smaller the black hole is, the higher the effective temperature and vice-versa.

The derivation of this effective temperature comes from the boundary conditions of the horizon. Assuming an observer just outside the horizon, this observer has to have an acceleration in order to not fall into the black hole. This accelerated frame of reference will see a thermal radiation by the Unruh effect \cite{PhysRevD.7.2850, 1975JPhA....8..609D, 1976PhRvD..14..870U}, where acceleration results in a thermal bath of particles. However, this accelerated frame of reference is at local equilibrium with the horizon of the black hole, which means the horizon shares that same thermal bath. This results in that observer seeing the black hole horizon at following temperature:
\begin{align}
	T_{obs} = \frac{ \hbar c^3}{4\pi G \sigma \sqrt{2M_{bh}r\left(1-\frac{2M_{bh}}{r} \right)} } \;, 
	\label{eq:Tobs}
\end{align}
where r is the position of the observer, $\sigma$ the Stefan-Boltzmann constant, $\hbar$ is the reduced Planck constant, c the speed of light and G the gravitational constant. This comes straight from the black hole's metric, which is assumed to be a standard Schwarzschild black hole, and the observers position. 

This temperature the horizon observer sees can then be redshifted to infinity, to obtain how a far off inertial observer like one from Earth would see the temperature in this horizon:
\begin{align}
T_{bh} = \frac{\hbar c^3}{8\pi GM_{bh}\sigma} \sim 10^{-7} K \left(\frac{\msun}{M} \right) \;.
	\label{eq:HawkT}
\end{align}

Of note is that this is for a non-rotating (0 spin) non-charged black hole. While charged black holes are not expected to exist, as their charge will simply attract opposite charges until the black holes becomes neutral again, rotating black holes are expected in both stellar black holes and some versions of PBHs as detailed in section \ref{sec:Form}.

From the temperature the emission can be estimated, which can only come from the black hole's own energy. Therefore the black hole will evaporate when it emits all of its energy, the majority of that energy being the one stored in the black hole's own mass. For massive black holes this is extremely unlikely, not only because their temperature is much lower, but because their mass means they are constantly accreting matter unless completely isolated. For smaller black holes however, their emission might surpass their accretion, especially when not in dense environments like the usual void between stars.

This permits us to define the lifetime of a black hole as the time it will take for it to evaporate through Hawking radiation. This is only an approximation, because it does not take into account accretion. However, because as the mass goes lower it evaporates even faster and the accretion lowers, it is very hard for a black hole with a non-ridiculous lifetime (i.e. smaller than the lifetime of the universe) to jump out of it purely through accretion. A good estimate of the lifetime is \cite{doi:10.1119/1.1571268}:

\begin{align}
	\tau(M) \sim 10^{64} \rm yr \left(\frac{M}{\msun}\right)^3 \;.
	\label{eq:lifetime}
\end{align}

The lifetime of PBHs sets the strongest constraint, as they can not be the DM if they do not exist after all. Following from above, only PBHs bigger than $10^{15} \rm g$ would survive to today. Still, we must be remember that, while expected, Hawking radiation is currently unconfirmed, and complete PBH evaporation is still an open question due to the information loss paradox \cite{1976PhRvD..14.2460H}. For example, \cite{2018arXiv181002336D} argued that the "Memory burden" could slow or stop the black hole evaporation. 

The model used to obtain the lifetime is also relevant. As mentioned, an isolated PBH is commonly assumed, but as PBHs form just after inflation they will interact with the radiation dominated, almost homogeneous early universe. \cite{2021arXiv211202818S} found that accounting for those interactions pushed the minimum mass of PBHs to $10^{14} \rm g$ instead, with \cite{2021arXiv210506504G} reaching a similar conclusion via a very different method of constantly accreting matter instead.

Another example of how variable the constraints obtained via PBH evaporation are can be seen by using a different metric to compute the Hawking radiation. \cite{2021arXiv210302815P} used the Thakurta metric to find very different results with Hawking radiation being notably boosted, though note that the metric they chose as an alternative is contentious as being a good description of PBHs \cite{2021EPJC...81..999H}, so their constraints are not reflected in figure \ref{fig:PBH_bounds}.

Therefore only using $\tau$ as a constraint can be problematic. Fortunately, PBHs of very small mass that survived to today would still emit a fair bit of Hawking radiation. PBH of mass $10^{15}$ to $10^{17}$ g would have high enough effective Temperature that we would expect them to emit quite a lot of high energy emission that would be completely isotropic from our point of view. If those PBHs are all of the DM, they would bump the extra-galactic background light up by a noticeable amount at high frequencies, but there is no such bump in current observations \cite{2020PhLB..80835624B, 2021PhRvL.126q1101C}. This makes the robust constraints which can be seen in figure \ref{fig:PBH_bounds}. 

There are hopes for even tighter constraints in the future by improving our detectors\cite{2021PhRvL.126q1101C}, but there are vastly diminishing results. In particular, even if we had perfect detectors Hawking radiation could only directly constrain PBHs below $10^{20} \rm g$ \cite{2022arXiv220101265A}, still leaving a good half of the asteroid mass range unconstrained. Another small issue that becomes relevant when the constraints are less stringent is that effect of spin now is non-negligible. As mentioned before the effective temperature, and so the emission and its spectra, of the PBH depend on its rotation, with highly rotating Kerr BHs emitting more. This change is expected to be less than an order of magnitude on most constraints, and so can be mostly neglected in mass ranges firmly ruled out by PBH evaporation. However the mass ranges with lower constraints become more dubious, as regions within the window but close to the constraints might be covered or on the opposite way ranges we think are currently impossible might be where PBHs are hidden. Constraints accounting for the spin are present in the literature \cite{2020PhRvD.101b3010A}, but spin becomes another factor that should be taken with care.

There might be even more factors however, as one of the more curious byproducts of this evaporation could be Planck relics. They are theorized particles of Planck mass ($\sim 10^{-6} \rm g$) left after a black hole evaporates \cite{1987Natur.329..308M}. They would be completely new physics, resulting from quantum gravity, and their existence is contentious. Nonetheless, even if they existed it is dubious if they could be all of the DM\cite{2021PhRvD.103d3532B}. They would need a massive quantity of PBHs with very small mass to evaporate in the early universe, and the increased emission and entropy would affect primordial nucleosynthesis. Exotic models exist within loop quantum gravity that could explain the lack of this effect via the relics forming before the big bang, which would be a big bounce in this particular case. If we were living in that specific universe with Planck relics as all of DM, it is likely we would not be able to detect them in any way in the close future due to their extremely low mass and unknown properties. Something to not is that the scenario described is not the same as the one where the DM are particles formed from the evaporation of PBHs \cite{2014PhRvD..89j3501F}, which is still possible but not described in this review as the DM would then require a new theoretical particle outside the Standard Model.

\subsection{Gravitational waves detectors} \label{sec:GW}

With the recent detection of gravitational waves (GWs) from a binary merger \cite{2016PhRvL.116f1102A}, LIGO opens a new window in astrophysics, the possibility of using GWs to detect black holes. While a substantial number of gravitational effects carry associated emission of GWs, the singular events that emit most significant bursts of GW in the frequencies LIGO can detect are binary mergers \cite{1993PhRvD..47.2198F}: 2 massive objects orbiting one another that slowly inspiral until they end up colliding and fusing into a single object.

Gravitational waves are not only emitted in the merger itself. Rather, they are the reason the merger itself happens. As two massive objects orbit one another they will emit gravitational waves from their movement, their moving gravitational potential perturbing the local space-time \cite{1964PhRv..136.1224P}. These are very small and initially undetectable, but similar to the Hawking radiation just the emission of GWs costs energy to the source. In the cases of the binaries the energy stored in their orbit is slowly emitted as GWs, resulting in the orbit becoming due to the energy loss. As the orbit decays however the tidal forces coming from the other object of the binary further perturb the source further, causing more emission of GWs and further decay. The same happens in reverse for the other member of the binary. In a perfectly symmetrical case the inspiralling binaries would meet in the middle, colliding and merging despite being otherwise completely isolated. The GWs emission is what turns the initially stable binary into a future merger.

The frequency of the GWs emission is in turn related to the orbits of the binary, rising as the orbits get shorter and peaking at the merger. A good benchmark is therefore the Innermost Stable Circular Orbit (ISCO), the last stable orbit before the merger. For a binary with masses $M_1$ and $M_2$, the frequency of the GWs from the ISCO is \cite{2022arXiv220502153F}:

\begin{align}
	f_{ISCO} \approx 4.4 \cdot 10^3 \; \rm Hz  
	\left(\frac{\msun}{M_1 + M_2} \right) \;,
	\label{eq:freISCO}
\end{align}

While the GWs emission peaks during the merger, LIGO only has enough resolution to see the waves from just before the merger \cite{1992Sci...256..325A, 2015CQGra..32g4001L}, the time scale of the GW emission at the beginning of the inspiral too large and the time scale during the merger too short. This works for using ISCO as an estimate though. From equation \ref{eq:freISCO}, knowing LIGO's frequency range of $\sim 10-10^4 $ Hz gives total mass ranges from 1 to $100 \msun$, in line with detections of current mergers. 

The GWs will have an amplitude, which is related to the characteristic strain $h_c$ . For a typical merger of an equal mass $m_{PBH}$ binary at a distance $d_l$, $h_c$ is approximately \cite{2022arXiv220502153F}:

\begin{align}
	\left| h_{c} (f) \right| \approx 4.54 \cdot 10^{-28}
	\left(\frac{m_{PBH}}{10^{-12}\msun} \right)^{5/6} \left(\frac{d_l}{\rm kpc} \right)^{-1} \left(\frac{f}{\rm GHz} \right)^{-1/6} \;,
	\label{eq:Hc}
\end{align}

For LIGO's sensitivity band the limiting characteristic strain is $\sim 10^{-22}$, though note that the timescale of the signal is also important as mentioned before. Besides the frequency range, the biggest limiting factor is distance.

\begin{figure}
	\includegraphics[width=\columnwidth]{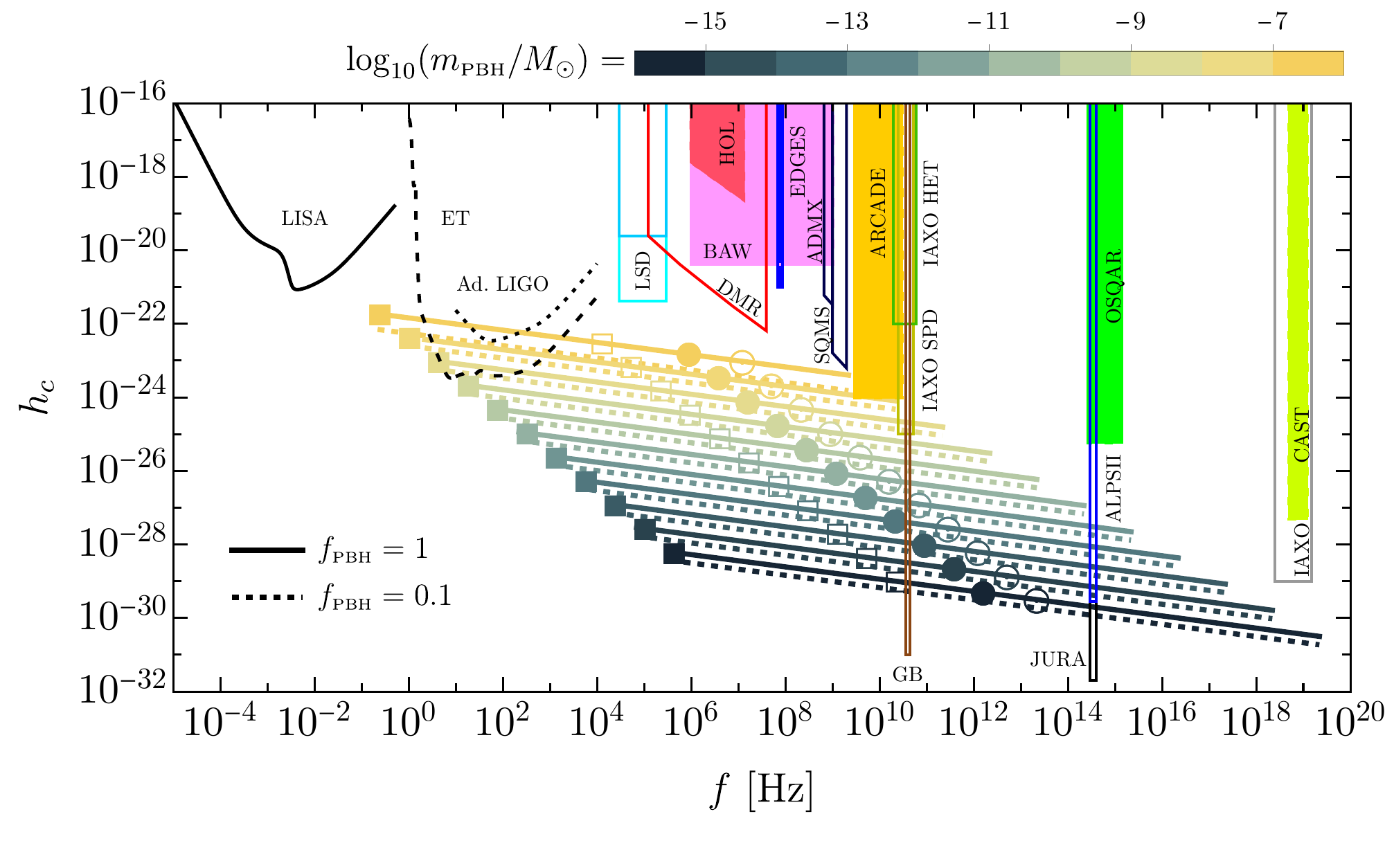}
	\caption{Figure showcasing the sensitivity band of current and future GWs detectors together with projections for the GW emission for small PBHs. Taken from \cite{2022arXiv220502153F}. Solid (dashed) lines indicate the $h_c$ spanned by an inspiralling binary with component masses $10^{-16}<m_{PBH}<10^{-6} \msun $ for $f_{PBH} = 1 \, (f_{PBH} = 0.1)$. The distance is taken as the minimum from which there will be a rate of 1 merger event per year. The signal ends at the ISCO, while the filled square, empty square, filled and empty circles all signal the time left until the merger, being $ \Delta t = 10$ Gyr, $\, \Delta t = 1$ day, $\, \Delta t = 1$ sec and $\Delta t = 10^{-3}$ sec, respectively. Main observatories shown are LIGO \cite{2015CQGra..32g4001L}, LISA \cite{2019arXiv190706482B} and the Einstein Telescope \cite{2010CQGra..27s4002P}, but high frequency observatories from \cite{2021LRR....24....4A} are also shown. A final remark is that this assumes there is no limitation in observation time, which is extremely optimistic for the higher frequencies. For more details I refer to \cite{2022arXiv220502153F}.}
	\label{fig:GWs}
\end{figure}

An estimation of the sensitivity band for current and future observatories is showcased in figure \ref{fig:GWs}, from \cite{2022arXiv220502153F}. I must repeat their warning that the figure assumes the observation time is only set by the frequency evolution of the inspirals, when time resolution is rather important factor (for more details see \cite{2022arXiv220502153F}), but the figure is very illustrative nonetheless.

Once detected the GWs can give individual information from the objects in the system. If the merger comes from a  binary of neutron stars (NS), then the GW emission can be used for example to constrain the neutron star equation of state \cite{2018PhRvL.121p1101A}. If the merger is of a BH binary then in theory their mass and spin can be derived from their GW emission. There is an important caveat however: while it can get some information of the individual members of the binary, a lot of it is degenerate with other parameters. It is also still very hard to distinguish between a NS and a PBH in any individual merger without also an electromagnetic counterpart. Currently for LIGO the standard for assigning wheres the GW event comes from a neutron star or a BH is based purely on mass. Without a visible counterpart LIGO cannot distinguish between a NS or PBH with NS like mass \cite{2020PhRvL.124g1101T}.

For the case of the spin, LIGO can obtain what is called as the effective spin of a binary BH \cite{2017Natur.548..426F}:

\begin{align}
	\chi_{eff} \equiv \frac{\left(m_1 \chi_1 + m_2 \chi_2 \right)}{M} = \frac{c}{GM} \left(\frac{\vec{S}_1}{m_1}+\frac{\vec{S}_2}{m_2} \right) \;,
	\label{eq:effspin}
\end{align}
where $\chi_i$ is a dimensionless spin parameter project over the z axis, but which can be related to the angular momentum spin of the star $\vec{S}_i$ as shown in the rightmost term above. $m_i$ is the mass of each object and M the total mass of the merger.

As can be seen, the effective spin is degenerate with individual value of each spin. The individual spin of one of the members of the binary can also be negative from the two BHs rotating in opposite directions. Thus, $\chi_{eff} \sim 0$ could mean low spins in both BHs or a very highly counter-rotating duo. An average $\chi_{eff} = 0$ in many mergers would simply point to the spin of each member of the binary not correlating with each other. 

Currently LIGO has showed the GW events it detects have a preference for a smaller effective spin \cite{2021arXiv211103634T}, very close to averaging 0, stronger than expected \cite{Safarzadeh_2020}. While it might seem encouraging for PBHs, which are expected to have spin close to 0 in most cases, the uncertainty is still too high and there have been individual events with indications of at least one of the members of the binaries having non-negligible spìn value \cite{2021PhRvL.126q1103B}. For now results on the spin are ambiguous, more robust statistics are needed. 

Another issue can be the current sensitivity of LIGO. As seen in equations \ref{eq:freISCO} and \ref{eq:Hc} comparing with figure \ref{fig:GWs}, the observatory probes only GW events from objects of the order of solar mass or higher and which also must be fairly close (or belong to niche subset which amplifies the signal, like being ultra-compact binaries \cite{2018PhRvL.121w1103A}). While any discovery of GW is greatly helpful, being a very novel field, LIGO itself is perhaps of limited usefulness for PBHs as DM. 

Current constraints in the mass range of LIGO exist from the merger rate. If PBHs were the DM there would be so many PBH binaries the merger rate would be much higher than the one we observe, even if we could not distinguish the PBHs from the stellar origin BHs. Take note however that the modelling of the merger rate is controversial, with perturbation of binaries by third objects lowering the number of mergers expected. If PBHs follow the typical Poissonian distribution the constraints should still be strong enough to exclude PBHs as DM for the $1-100 \msun$ mass range \cite{2020PhRvD.101d3015V, 2021PhRvD.103b3026W}, but results vary depending on how the PBHs clusters are handled, with LIGO constraints not enough to rule PBHs out as all of the DM in \cite{2020JCAP...09..022J,2021PhRvL.126e1302J}.

There are still other ways in which GW detectors can be helpful however. An option to help distinguish PBHs are more statistical arguments: for stellar origin black holes we expect their number of mergers to peak with star formation, as they form from stars that are very massive with very short lives. This peak would be situated at $z\sim 1$ \cite{2016PhRvL.116m1102A}, while in contrast PBHs would start merging at much higher redshifts and the number of mergers would increase monotonically with redshift. Unfortunately, current detectors are incapable of distinguishing stellar origin from non-DM PBHs this way. Even for the most massive binaries LIGO does not have enough resolution to resolve mergers at redshift 1\cite{PhysRevLett.116.101102}, so trying to see where mergers would peak in a redshift distribution is not viable.

The solution is to wait for more powerful next generation detectors \cite{2021ApJ...912...53W}, particularly LISA \cite{2019arXiv190706482B}, but even with the improved resolution there is no guarantee of much better odds looking at figure \ref{fig:GWs}, particularly if we are searching for very light PBHs in the planetary or asteroid mass range.

An interesting alternative is not searching for individual events, but for their added effect. Just how like multiple sources we cannot identify individually contribute to a background in radiation, the same applies for GWs. Mergers we do not have resolution to identify will still emit GWs, and joined will create a non-zero background for our detectors. This incoherent superposition of unresolved signals is commonly identified as the Stochastic Gravitational Wave Background (SGWB). 

The frequency range of the SGWB is expected to be the same as the unresolved mergers that form it but redshifted, and the characteristic strain can be estimated as:

\begin{align}
 h_{c} (f) \approx 2 \cdot 10^{-31} \left(\frac{\Omega_{GW} (f)}{10^{-7}} \right)^{-1} \left(\frac{f}{\rm GHz} \right)^{-1} \;,
	\label{eq:HcSWGB}
\end{align}
where $\Omega_{GW}$ is the energy density of the spectra at a given $f$.

Just from the level of SGWB we would expect to see on LIGO some constraints can already be set for solar mass ranges \cite{2017JCAP...09..037R, 2017PDU....18..105C}, while the prospects for future detectors are also known \cite{2016PhRvL.117t1102M,2019ApJ...871...97C}.

A significant problem can also be correctly identifying the SGWB. There will also be quite a number of mergers from stellar black holes and other massive objects, so simply finding a signal is no smoking gun for PBHs. However, while it is extremely hard to obtain individual values from the background (it is the addition of a lot of unidentified mergers after all), the background follows the same statistics as the mergers. This means if the SGWB is dominated by signals from PBHs, which we expect if they are all of the DM, we would see effects coming from the different redshift distribution mentioned above, which could be used to put improved constraint on subsolar PBHs in the future\cite{2022MNRAS.510.6218M}. Another possibility is to see if the anisotropy of the SGWB mirrors that of the DM distribution which could also create strong constraints for solar mass PBHs \cite{2021arXiv210701935W}.

The SGWB can have an origin different than unidentified mergers though. As it just appears as a background in our detectors, any isotropic GW signal that completely covers our sky could be part of it. As GWs are quite literally aftershocks in space-time, we expect any event which significantly alters the flat space time metric to have some GW associated to them. This does not extend only to massive objects, but also to perturbations, like the ones creating the PBHs. The same primordial curvature perturbations that create PBH will at second order originate scalar-induced gravitational waves \cite{PhysRevD.76.084019}. As the perturbations are expected to be isotropic, the emission would result in another background that could give information about the curvature perturbations, and so its PBH generation and abundance \cite{2021arXiv210812475K, 2021iSci...24j2860Y}. Of note is that while perturbations require physics outside the standard model to be produced in large enough quantities to create PBHs, this background would depend only on the perturbations themselves and thus requires no knowledge of the specific new physics to simulate. The frequency we expect would depend only on the patch within the horizon, being directly related to the mass within the horizon $m_h$ described in section \ref{sec:Form} \cite{2010PThPh.123..867S}:

\begin{align}
 f \approx 5 \rm kHz 
 \left(\frac{M_h}{10^{-24} \msun} \right)^{-1/2} \;.
	\label{eq:frec_SWGB}
\end{align}
Though that would be heavily redshifted today. Likewise, if the PBHs origin was unrelated to the collapse of primordial perturbations, then the SGWB associated to their formation may not exist. Note also that SGWBs may also be created by other plethora of early Universe GW production mechanisms, almost all of which would imply physics beyond the Standard Model \cite{2022arXiv220502153F}

While LIGO's limitations still apply \cite{2022A&A...660A..26B}, this makes the future LISA mission even more interesting. Particularly, the range of LISA would cover the GW background associated to perturbations that would result in $10^{-12} \msun$ PBHs \cite{2019PhRvL.122u1301B}. Considering this is one of the few windows remaining for PBHs to be all the dark matters, as can be seen in figure \ref{fig:PBH_bounds}, this is a very exciting possibility.

\subsection{Other constraints}

While PBHs of mass in the range of planets and asteroid presents a significant challenge to detect, this does not extend to masses higher than $10 \msun$. This is natural, as the only compact objects known above that mass are other black holes. While this makes distinguishing the primordial nature of the BH much harder, it also means such objects cannot hide among low luminosity objects like brown dwarves, planets or asteroids. Furthermore, the abundance of stellar origin black holes is itself closely limited by the number of baryons in the universe present in the primordial nucleosynthesis \cite{2016RvMP...88a5004C}. As the DM is expected to be about 5 orders of magnitude more common than baryonic matter, if such PBHs were all the DM BHs from primordial nature would significantly outnumber their stellar origin siblings.

The typical way to detect BHs is through their dynamical effects. By studying the motions of object within our galaxy it is possible to estimate the gravitational forces they are subject to, and from them the masses and motions of other close objects. This is for example how Sagittarius A*, the supermassive black hole within our own Milky Way, was initially detected \cite{1994RPPh...57..417G}. As the mass of the PBH rises, it will also rise its gravitational potential and so their effect on nearby objects. This is the basis for the dynamical constraints we see on figure \ref{fig:PBH_bounds}, coming from how a large number of massive PBHs would encounter and then disrupt wide binary star systems within our Milky Way \cite{2004ApJ...601..311Y, 1985ApJ...290...15B}.

A more atypical way to obtain further constraints is from the accretion effects such a large number of massive PBHs would bring. PBHs accrete matter like any other BH, and like any other BH as the matter falls into the event horizon it will emit a portion of its energy as radiation. The accretion due to gravity for a given object commonly follows the Bondi regime, with a squared dependence on mass $M$: 
\begin{align}
\Dot{M}_{\rm B} = \, & \frac{\pi G^2 M^2 \rho_s}{c_{s}^3} ~,
 \end{align}
where $c_s$ is the sound speed and $\rho_S$ is the density of the surrounding medium and .

For low mass PBHs, the effects of accretion into their surroundings is completely negligible, but at higher masses, particularly above $10 \msun$, accretion becomes very significant. Constraints can be made from both the direct emission we would expect to see from such accretion in X-rays \cite{2017PhRvL.118x1101G}, as well as by measuring how much PBHs would heat the surroundings with their accretion and if we see such effects in near dwarf galaxies \cite{2021ApJ...908L..23L}.

One interesting advantage of such type of constraints is that they only depend on the DM distribution and basic accretion physics. Formation path, spin and even the particulars of our cosmological model are not a major factor, and thus constraints can be made to be model independent without losing most of the strength of their bounds \cite{2022JCAP...03..017T}. The main limits on accretion constraints come from the reliability of the DM density distribution models for our Milky Way and other galaxies.

Currently, accretion constraints heavily limit the abundance of PBHs to be even a very small fraction of the DM from 10 to $10^7 \msun$ \cite{2021ApJ...908L..23L}.

\section{Non-monochromatic constraints} \label{sc:non-diff}

Having detailed the main sources of PBH constraints, one can perhaps wonder why monochromatic constraints have remained so popular, why extended mass distributions of PBH are not the standard. A priori, adding multiple masses should likewise end up adding multiple constraints in a fairly simple way, as can be seen directly with equations (\ref{eq:A_m}) and (\ref{eq:lifetime}). Those mass ranges are still considered constrained to an extended mass distribution in our previous figure, as the extended mass distribution should be roughly equivalent to monochromatic mass with slightly lower abundance, but using an extended mass distribution would be both more accurate and likely increase the strength of the constraints, as constraints in adjacent mass ranges could contribute too.

Unfortunately, while the cases for microlensing (in the simplest form) and PBH evaporation work very well, this is because their constraints come from effects that have a direct relation with mass. Both equations ($\ref{eq:A_m}$) and ($\ref{eq:lifetime}$) are analytic expressions that contain the mass (through the Einstein angle  of equation (\ref{eq:Einst_angl}) in the microlensing case). In other constraints the effects can be more indirect, a good example being gravitational waves: we would not only see the PBHs of a certain mass, but their abundance and spread would directly affect how many mergers we expect to see if they are the dark matter, and the difference in masses between binaries has a pretty direct effect on the frequency and shape of the GW burst their merger would create. Another good example can be simulations of dynamic effects the PBHs would have. While technically possible to account for an extended mass distribution, as multiple bodies simulations their relation is highly non-linear, so the time and computational cost can increase exponentially. Even in cases different to those two, if no direct analytical relation exist between the constraint and mass it is much harder to account for the extended mass distribution, as it would require numerically integrating over different masses all the process rather than just one equation. 

\cite{2018JCAP...01..004B} studied the issue in a thorough way, and reached a similar conclusion. To convert a monochromatic constraint into one for an extended mass function you would need to compute a function g:

\begin{align}
	g(M_{eq}, p_j)= \int dM \frac{d \Phi_{EMD}}{dM}g(M, p_j) \;.
\end{align}
 $g$ is a function that encloses the details of the underlying physics of the particular constraint and depends on the PBH mass M and a set of astrophysical parameters $p_j$. $\Phi_{EMD}$ is the shape of the extended mass distribution function. $M_{eq}$ is the main result, the equivalent mass to move from a monochromatic constraint to an extended distribution one, and it is defined as the individual mass in the monochromatic constraint that the extended function is effectively equivalent to. If an extended distribution has a $M_{eq} = 10^{-6} \msun$, it would constrain the abundance of PBHs of the extended distribution to the same degree as $M=10^{-6} \msun$ is in the monochromatic constraints.

$g$ must be determined individually for each different constraint, and should enclose both the physics of the constraints and the way the observations were taken (for more details I refer to \cite{2018JCAP...01..004B}). Again, for some constraints $g$ will be an analytic expression we know and there will not be any major issues resolving the integral above and finding $M_{eq}$. But for other constraints g is much, much harder to determine

Something else to note is that while this allows us to find constraints for extended mass distributions through only the monochromatic constraint, this does not allow us to combine 2 different set of constraints. It might be tempting to consider that 2 nearby constraints can together constrain the same extended mass distribution in higher degree than each alone. However, the physics between the 2 different sets of constraints differ, or in other words, the $g$ above is different. The equivalent mass of the extended distribution will have to be computed individually, and the more stringent constraint taken. For purely monochromatic constraints this much clearer, as there is no addition of different types of constraints, the bigger one is the one we apply, simple as that. But the same applies to our extended extended distributions, as we are converting our constraints from monochromatic ones.   

Besides the one described above, other methods exist\cite{2017PhRvD..96b3514C} to translate monochromatic constraints into extended ones, but have similar characteristics and difficulties.

The type of extended function expected is also open to debate. A lognormal mass function is the one commonly believed to be a good approximation for the standard PBH formation path of fluctuation collapse\cite{2016PhRvD..94f3530G}:

\begin{align}
	\psi(M) = \frac{f_{PBH}}{\sqrt{2\pi} \sigma M} exp\left(-\frac{log^2(M/M_c)}{2\sigma^2} \right) \;,
	\label{eq:lognormal}
\end{align}
where $M_c$ is the mass of the peak, the one commonly associated with the monochromatic constraint, and $\sigma$ is the width of the distribution. The value of these parameters depends directly of the model used. However, as keeps repeating, particular models can result in particular mass functions, including power functions and even very sharp mass distributions that are very close to a purely monochromatic one.

Of note is that the extended mass distribution can, and in many cases will, change with time. Mergers between  PBHs and accretion can bring major changes by significantly raising the abundance of higher mass PBHs \cite{2020PhRvD.102d3505D}, while on the opposite end very small PBHs will slowly evaporate by a different rate depending on their own mass, changing the shape of the distribution \cite{2022arXiv220305743M}. While this may bring some slight changes between the formation mass distribution and the current one it does not affect most constraints, including microlensing, gravitational waves and evaporation, as they are based on observations of the nearby universe and so constrain the mass distribution at current redshift 0. Constraints based on high redshift, like those coming from the CMB, however will have to be shifted to what currently are higher masses.

Overall, extended mass distributions seem to create more stringent constraints than monochromatic ones, but it is not possible to generalize for all cases and graphs have to be individualized for each type of distribution. In comparison, the monochromatic constraints figure is generic and much easier to produce, and a sufficiently good estimate to merit its continued usage.

While different, something similar happens with clustered distributions. To begin with, the curvature fluctuations that originate the PBHs in our standard formation path are Poissonian \cite{2018JCAP...10..043B, 2018PhRvD..98l3533D, 2018PhRvL.121h1304A}, and so it is expected for a large number of small clusters of PBHs to form. However, we also expect most of those small clusters to have evaporated by now \cite{2003ApJ...594L..71A}, and to be too diffuse to affect the microlensing constraints for example \cite{2022arXiv220102521P}. 

Other formation models can result in cases where the PBHs form in a much more clustered distribution rather than the typical spread \cite{2019EPJC...79..246B}, and even in the typical fluctuation collapse the presence of non-gaussianities can bring forth non-negligible clustering \cite{2019PTEP.2019j3E02S, 2020JCAP...03..004Y}. Like before, the change in signals and constraints must be treated individually for each mass range in question. Care has to be given though, because while clustering can decrease the constraints for some mass ranges, on some cases it can also introduce new signals that should not be overlooked. A good example are binary mergers: clustering significantly decreases the constraints of LIGO, as most mergers would have occurred in the early universe due to PBHs being more concentrated, but those previous mergers also create a very significant stochastic GW background \cite{2020JCAP...11..036A} that would be within the current LIGO/VIRGO sensitivity \cite{PhysRevD.104.022004}.

Another example can be with microlensing. While very strongly clustered PBHs are expected to form a single much larger lens rather one per PBH, which should a priori reduce the number of lenses and so of events we expect to see, the number of events expected can actually raise if we have one of the PBH clusters close enough by chance \cite{2022arXiv220304209G}.

Naively we would expect PBH evaporation and accretion constraints to not be strongly affected by clustering, but understanding of both the physical reasoning of the constraints and also the particularities of the clustered distribution are needed before applying monochromatic non-clustered constraints.

\section{Existing windows} \label{sc:window}

Among all the discussed constraints for smaller PBHs that could be all the dark matter, there still remains a sizable window between $10^{-16}$ and $10^{-11} \msun$. This window is extremely hard to probe with usual methods, as they are PBHs too small to be detected either through gravitational waves emission or microlensing events by current instruments, but also are too big to be affected by evaporation via Hawking radiation. Their mass, roughly corresponding to the one of an asteroid, also makes it exceedingly hard to detect them through conventional methods in big enough numbers to say anything about their abundance, even if somehow a few could be found amid countless opaque asteroids and similar objects.

There have been a number of attempts at constraining the mass range still, most trying to use the effect such PBHs would have onto more standard objects like stars and predict what observational consequences it could leave.

A novel approach based on these ideas was proposed in \cite{2009arXiv0901.1093R}, where they expose that asteroid mass PBHs could randomly traverse through a star and then fall to the core of said star via energy loss coming from dynamical friction. If the DM contains PBHs, these would be present in the cool molecular gas clouds in which star formation happens and, in cases where the relative velocities are low enough, the PBHs would undergo adiabatic contraction as the baryonic mass in the form of gas cools down during the formation of a protostar, resulting in a number of PBHs bound to the newly born star. A portion of the PBHs orbiting close or within the star would eventually lose energy via dynamical friction and fall to the core of the star, thus being captured. Despite the small mass of the PBHs, once settled in the stellar core they would start accreting, growing up to a total mass that might be a large fraction of the total mass of the star.

This raises the possibility of black hole with a typical mass of a star but below the Chandrasekhar mass existing, which cannot be explained in ordinary stellar evolution but is much easier to detect. If any such black hole could be detected today, it would therefore point to a possible origin in a PBH and thanks to forming via such indirect methods it would avoid abundance constraints in the solar mass range. Similarly, the lack of such objects could be used to place constraints on the asteroid mass range. However, \cite{2009ApJ...705..659A} questioned that this process could occur in the Milky Way, finding the rate of this PBH capture to be negligible because capture by dynamical friction is very slow and the current DM density too low.

The possibility of PBH capture in the present Universe was reexamined by \cite{2013PhRvD..87b3507C}, considering star formation in globular clusters formed in dense dark matter halos made of PBHs in the asteroid mass range. If the star become a neutron star while it had a PBH within itself, the much higher densities of the object would result in an almost instant accretion by the PBH. With high enough DM densities, the process is common enough as to imply no neutron star would survive, making constraints based on the detection of such neutron stars possible. While energy loss via dynamical friction was still very slow, high DM density meant rare very eccentric orbits or orbits wholly within the star could be populated by PBHs, and only one PBH was required to be captured for the star to be accreted. The impact of eccentric orbits to the capture rate was considered in a follow up study \cite{2014PhRvD..90h3507C}, which found an enhanced capture rate that made the constraints tighter around the same mass range. Both works however considered globular clusters with a very high DM and baryonic density, whereas in fact globular clusters may form from gas cloud fragmentation without involving any DM, so their constraints are not readily applicable \cite{2019JCAP...08..031M} and so do not appear in figure \ref{fig:PBH_bounds}.

A further refinement of this method of PBH capture is present in \cite{2022arXiv220513003O}, which concluded that the higher DM density and lower velocity dispersion in DM haloes at high redshift ($z \sim 20$) would allow PBHs to be captured by common low mass main sequence stars. While no bounds based on neutron star survival can be made, as most captures happened at very high redshift, a substantial number of sub-Chandrasekhar mass BHs would exist as results of such captures if PBHs in the asteroid mass range are all of the DM.

\cite{2014JCAP...06..026P} presented a similar case of capture of PBHs but by neutron stars directly. Rather than dynamical friction, the energy loss of the PBH was via tidal deformations of the neutron stars, which was supposedly more efficient and thus would end up with neutron stars capturing the PBHs much more easily, severely constraining the mass range as DM candidates. \cite{2014arXiv1402.4671C, 2014PhRvD..90j3522D} however cast doubts on it, arguing both tidal deformations and dynamical friction were roughly equivalent, and models agreed with previous dynamical friction calculations. 

Overall, despite the interesting approaches, asteroid mass range PBHs remain the last big windows where PBHs could provide all the dark matter. There exists more exotic PBH formation models that avoid most of the other constraints, but if using standard inflation perturbation collapse with as few extra elements as possible they remain perhaps the last possibility to explain the DM. In the future LISA could constrain this simple case through the SGWBs the PBH formation would produce \cite{2019PhRvL.122u1301B}, but today it still remains one of the most interesting windows in the PBH mass range.

\section{Possible detections}\label{sec:Direct detection}

Most constraints on PBh abundance are based on assuming a population of PBH with a certain abundance and mass range, studying a possible observable consequence their existence would bring and then using the lack of said detections to set constraints. The inverse however is possible, that is, to explain an observed but unexplained event through a population of PBHs.

The most famous example of this is the one previously cited, where a population of PBHs could explain the LIGO GWs merger events observed \cite{2016PhRvL.116t1301B}. While currently the mergers can all be explained through only stellar origin BHs, models suggest that a combined population of PBHs and stellar origin BHs mergers is the one that fits best the rate observed by LIGO \cite{2021JCAP...03..068H, 2021JCAP...05..003D}, though with an abundance far below the one needed for PBHs to be the dark matter and with uncertainties in astrophysical and primordial formation models not allowing for an unequivocal answer \cite{2022PhRvD.105h3526F}. Exotic variations also exist where PBHs outside the LIGO mass range are not all of the DM and end up creating the mergers seen by LIGO \cite{2021PDU....3200833E,2020PhRvL.125r1304K}.

Another good example are microlensing events. Famously, while the MACHO experiments constrained compact objects to not be all of the DM they also found a higher number of microlensing events than expected by objects in the disk \cite{2000ApJ...542..281A}. Microlensing surveys like HSC today also find a number of microlensing events caused by unknown compact objects \cite{2019NatAs...3..524N} that could be easily be explained by PBHs, though it has to be stressed than in a much lower abundance than the one needed for them to be all of the DM. Finally, \cite{2022MNRAS.512.5706H} argued that PBHs as all of the DM could explain the microlensing effects seen on the quasars, but their results contradict other microlensing studies of quasars \cite{2022arXiv220304777E}.

Another event that has commonly been explained as a possible result of PBHs is the excess of 511 keV photons from the center of the Milky Way relative to expectations found by telescopes SPI/INTEGRAL/COSI \cite{2005MNRAS.357.1377C, 2020ApJ...895...44K}. Very small PBHs could be the source of the increased radiation via their Hawking emission \cite{2021PhRvD.104f3033K}, though it can also be explained by other conventional means or particle DM \cite{2021arXiv210914955M}, and even if the origin were PBHs they would only make up a small fraction of the DM.

Perhaps the most famous and current case are the latest finds by NANOGrav \cite{2020ApJ...905L..34A}. Within a pulsar-timing data set comprising 12.5 years there was evidence of a stochastic common-spectrum process. The signal was not found to have a quadrupole correlation that would confirm the origin of such process as being from a SGWB, but it is still a popular and reasonable explanation for the process.

In particular, a SGWB with parameters that could result in the process observed in NANOGrav could in turn have the origin in PBH formation \cite{2021PhRvL.126d1303D, 2020arXiv201003976D}. As mentioned in section \ref{sec:GW}, the formation of PBHs by perturbation collapse generates its own SGWB. For the NANOgrav case, the formation of a small abundance of subsolar mass PBHs would create a SGWB capable of generating the stochastic process present in the data \cite{2021PhLB..81336040K}.

It is important however to stress that the abundance of PBHs that would be generated is far below the one needed to explain all of the DM if only using subsolar mass PBHs, at least as long as subsolar mass PBHs are the only ones produced. There exists the possibility of a very broad PBH mass range distribution, where a small fraction of subsolar mass PBHs would be generated as the end point of a distribution with a peak around the asteroid mass range \cite{2021PhRvL.126d1303D,2021PhLB..81436097S}, which could explain all of the DM. This extremely broad distribution means there would be a non-negligible abundance of PBH in the microlensing range too, which the HSC would probe in the future \cite{2021PhLB..81436097S}.

Overall, direct detection of PBHs remain a tantalizing if challenging prospect. While several events exists which could be explained by their presence, all of them would require them to be only a small fraction of the DM, and several, like the galaxy center excess, have even been used to create constraints.

\section{Conclusions and future prospects}\label{sec:future_perspectives}

As of today, only PBHs within a small range of masses between $10^{-16}-10^{-11} \msun$ could make up all of the DM, assuming they form with the smallest deviation possible to the standard model. But while relatively small this range of masses remains challenging to constrain, comprising of PBHs too small to distinguish from diffuse matter and too massive to emit significant amounts of Hawking radiation.

In the years following this review it is expected for plenty of experiments and new instruments to help further constraint the abundance of PBHs \cite{2021JPhG...48d3001G, 2022arXiv220308967B}. While helpful, most of these tighter constraints are in already constrained mass ranges, so they offer gradual improvement rather than resolving the remaining mass window.

There are two possible exceptions. The first are the so called 3rd generation GW instruments. On the ground the Cosmic Explorer \cite{2017CQGra..34d4001A,2019BAAS...51g..35R} and Einstein Telescope \cite{2010CQGra..27s4002P,2020JCAP...03..050M} could, if approved and build, be able to observe binaries with total mass of $\sim 10 \msun$ up to redshifts $z \sim 100$. While not on the most interesting mass ranges, the possibility of detecting mergers at extremely large redshifts, before the first stars, remain one of the few ways in which PBHs existence could be undeniably proved or harsher constrains be made \cite{2022arXiv220411864N, 2021JCAP...11..039D}. Furthermore, if PBHs are all of the DM in the asteroid mass range, they would be captured and accrete a good portion of the lower mass longer lived first stars \cite{2022arXiv220513003O}, so even the asteroid mass range could be probed indirectly by looking for sub-Chandrasekhar mass mergers.

On space, LISA \cite{2017arXiv170200786A} frequency range offers the possibility of detecting the SGWB emitted during the formation of asteroid mass PBH \cite{2019PhRvL.122u1301B}, as mentioned in section \ref{sec:GW}. While the emission is only expected in the case of PBH formation through the standard perturbation collapse, it remains one of the few ways to probe the case where PBH are all of the DM. LISA should also be able to confirm if the stochastic process detected by NANOGrav is related to PBH formation \cite{2021PhRvL.126e1303V}. While only tangentially related, next generation pulsar timing arrays like ngVLA \cite{2018arXiv181006594N} could also help with the detection of SGWBs, particularly if the NANOGrav signal persists.

Another exciting prospect are X-ray pulsars microlensing, which avoid finite body effects due to pulsar's small size. Current X-ray telescope AstroSat \cite{2014SPIE.9144E..1SS} could already create some constraints around the $10^{-14} \msun$ range with 300 days of observations \cite{2019PhRvD..99l3019B}. Future missions Athena \cite{2015JPhCS.610a2008B} and Lynx \cite{2018arXiv180909642T} could improve upon those. eXTP \cite{2016SPIE.9905E..1QZ} could widen the constraints to reach masses $10^{-15}-10^{-13} \msun$ \cite{2019PhRvD..99l3019B}, which would result in PBHs needing very narrow and fine tuned mass distributions to make up all of the DM.

Important amounts of clustering could relax the microlensing constraints, though by how much is still debated \cite{2022arXiv220304209G}. Nonetheless, relaxing microlensing constraints would widen the window where PBHs can make up all of the DM considerably and change future prospects. Still, in analogy to the case in \cite{2020JCAP...11..036A}, significant clustering would also result in much increased mergers at earlier redshifts that would leave a very relevant SGWBs. Future GW detectors like the already mentioned Einstein Telescope could make up some of the lost microlensing constraints in this case.

Overall, asteroid mass PBHs are today still a rather interesting candidate to be all of the DM. While much more constrained that when they first awoke interest after the LIGO detection, a window still remains, one which is very challenging to close with current instruments. Next generation experiments should be able to probe the window with both microlensing and GWs, either closing it or perhaps finally answering the decades long question that is the DM.

\funding{This work was supported in part by Spanish grant CEX-2019-000918-M funded by MCIN/AEI/10.13039/501100011033.}

\acknowledgments{I would like to thank discussions on the PBH topic with Albert Escriv\`a, Ivan Esteban, Cristiano Germani, Jordi Miralda, Jordi Salvad\'o and Javier G. Subils.
}





\abbreviations{PBHs: Primordial Black Holes\\ DM: Dark matter\\
BH: Black Hole\\
GW: Gravitational Wave\\
CMB: Cosmic Microwave Background\\
LIGO: Laser Interferometer Gravitational-Wave Observatory\\
SGWB: Stochastic Gravitational Wave Background\\
NANOGrav: North American Nanohertz Observatory for Gravitational Waves\\
ISCO: Innermost Stable Circular Orbit\\
}


\appendixtitles{no} 


\reftitle{References}


\externalbibliography{yes}
\bibliography{PBH_constraints.bib}


\end{document}